\documentclass[aps,reprint,letterpaper,amsmath,amssymb, showpacs, superscriptaddress, longbibliography]{revtex4-1}
\usepackage{graphicx}
\usepackage[absolute]{textpos}

\begin{document}
\begin{textblock*}{8.5in}(0.1in,0.25in)
\begin{center}
PHYSICAL REVIEW B \textbf{93}, 125302 (2016)
\end{center}
\end{textblock*}
\begin{textblock*}{2.5in}(5.06in,4.50in)
{\small \doi{10.1103/PhysRevB.93.125302}}
\end{textblock*}

\title{Large anisotropy of electron and hole $g$ factors in infrared-emitting InAs/InAlGaAs
self-assembled quantum dots}

\author{V. V. Belykh}
\email[]{vasilii.belykh@tu-dortmund.de}
\affiliation{Experimentelle Physik 2, Technische Universit\"{a}t Dortmund, D-44221 Dortmund, Germany}
\author{D. R. Yakovlev}
\affiliation{Experimentelle Physik 2, Technische Universit\"{a}t Dortmund, D-44221 Dortmund, Germany} \affiliation{Ioffe
Institute, Russian Academy of Sciences, 194021 St. Petersburg, Russia}
\author{J. J. Schindler}
\affiliation{Experimentelle Physik 2, Technische Universit\"{a}t Dortmund, D-44221 Dortmund, Germany}
\author{E. A. Zhukov}
\affiliation{Experimentelle Physik 2, Technische Universit\"{a}t Dortmund, D-44221 Dortmund, Germany}
\author{M. A. Semina}
\affiliation{Ioffe Institute, Russian Academy of Sciences, 194021 St. Petersburg, Russia}
\author{M. Yacob}
\affiliation{Institute of Nanostructure Technologies and Analytics (INA), CINSaT, University of Kassel, Heinrich-Plett-Str. 40, D-34132 Kassel, Germany}
\author{J. P. Reithmaier}
\affiliation{Institute of Nanostructure Technologies and Analytics (INA), CINSaT, University of Kassel, Heinrich-Plett-Str. 40, D-34132 Kassel, Germany}
\author{M. Benyoucef}
\affiliation{Institute of Nanostructure Technologies and Analytics (INA), CINSaT, University of Kassel, Heinrich-Plett-Str. 40, D-34132 Kassel, Germany}
\author{M. Bayer}
\affiliation{Experimentelle Physik 2, Technische Universit\"{a}t Dortmund, D-44221 Dortmund, Germany}
\affiliation{Ioffe Institute, Russian Academy of Sciences, 194021 St. Petersburg, Russia}

\received{7 December 2015} \revised{11 February 2016} \published{9 March 2016}

\begin{abstract}
A detailed study of the $g$-factor anisotropy of electrons and holes
in InAs/In$_{0.53}$Al$_{0.24}$Ga$_{0.23}$As self-assembled quantum
dots emitting in the telecom spectral range of $1.5-1.6$~$\mu$m
(around 0.8~eV photon energy) is performed by time-resolved
pump-probe ellipticity technique using a superconducting vector magnet. All
components of the $g$-factor tensors are measured, including their
spread in the quantum dot (QD) ensemble. Surprisingly, the electron $g$ factor
shows a large anisotropy changing from $g_{\mathrm{e},x}= -1.63$ to
$g_{\mathrm{e},z}= -2.52$ between directions perpendicular and
parallel to the dot growth axis, respectively, at an energy of
0.82~eV. The hole $g$-factor anisotropy at this energy is even
stronger: $|g_{\text{h},x}|= 0.64$ and $|g_{\text{h},z}|= 2.29$. On
the other hand, the in-plane anisotropies of electron and hole $g$
factors are small. The pronounced out-of-plane anisotropy is also
observed for the spread of the $g$ factors, determined from the spin
dephasing time. The hole longitudinal $g$ factors are
described with a theoretical model that allows us to estimate the QD
parameters. We find that the QD height-to-diameter ratio increases
while the indium composition decreases with increasing QD emission
energy.
\end{abstract}

\maketitle

\section{Introduction}

Carriers in semiconductors, electrons or holes, are quasiparticles,
whose effective spin is strongly modified by the motion in crystal
lattice. Consequently, carrier confinement in semiconductor quantum
wells (QWs) and quantum dots (QDs) significantly modifies the spin
properties. In particular, the QD confinement suppresses the
spin-orbit interaction \cite{Khaetskii2000}, resulting in long spin
coherence times \cite{Greilich2006Sci}. Furthermore, by tailoring
the QD potential profile it is possible to adjust electron and hole
$g$ factors, which characterize the susceptibility of the spins to a
magnetic field. These features make semiconductor QDs attractive
systems for manipulating carrier spins both for fundamental and
applied research \cite{Benson2008}.

The carrier confinement in semiconductor heterostructures also changes the
band gap energy, i.e., the energy between the lowest conduction band
and the highest valence band. This leads to a modification of the
\emph{electron} $g$ factor, which can be reasonably well described by the
Roth-Lax-Zwerdling relation \cite{Roth1959}, as demonstrated for
CdTe/(Cd,Mg)Te \cite{Sirenko1997} and GaAs/(Al,Ga)As QWs
\cite{Yugova2007Univ,Kiselev1998} despite its derivation for bulk
semiconductors. However, considerable deviations from this relation
were found recently in QDs with band gap energy smaller than 1.2~eV
\cite{Belykh2015}. Another consequence of the quantum confinement
and the related symmetry reduction in epitaxially grown QWs and QDs
is the appearance of the electron $g$-factor anisotropy,
characterized by the difference between the transverse (magnetic
field in the sample plane) and longitudinal (magnetic field parallel
to the growth axis) $g$ factors $\delta
g_\text{e}=g_{\text{e}\perp}-g_{\text{e}||}$. The difference is
induced by the modification of the hole states in the valence band
(e.g. the splitting of light-hole and heavy-hole states) and their
admixture to the electron states in the conduction band.  This
effect is quite small: the reported $|\delta g_\text{e}|$ values for
QWs do not exceed $0.1-0.2$
\cite{Hubner2011,Salis2001,Sirenko1997,Malinowski2000,Nefyodov2011,Nefyodov2011a,
Jeune1997}. Also for standard (In,Ga)As/GaAs QDs emitting at photon
energies $E>1.3$~eV the reported differences do not exceed $0.2$
\cite{Schwan2011,Yugova2007,Meyer2001}. Only for QDs emitting at
lower energies indications for a strong anisotropy of the electron
$g$ factor were found in electrical \cite{Alegre2006} and optical \cite{VanBree2016} measurements.
An in-plane electron $g$-factor anisotropy was also observed, while being rather small with the
$g$-factor difference not exceeding 0.05
\cite{Schwan2011,Yugova2007,Hapke-Wurst2002}. The \textit{hole} $g$
factor, on the other hand, is controlled by the complex spin level
structure of the valence band and can vary strongly and
nonmonotonically with changing quantum confinement. The out-of-plane
and in-plane anisotropy of the hole $g$ factor can be quite large
both for QWs and QDs \cite{Schwan2011,Crooker2010,Gradl2014,
Krizhanovskii2005,Koudinov2004,Zhukov2014}.

In this paper, we present a comprehensive investigation of the
$g$-factor anisotropy of electrons and holes in
InAs/In$_{0.53}$Al$_{0.24}$Ga$_{0.23}$As quantum dots emitting in
the telecom spectral range of $1.5-1.6$~$\mu$m (around 0.8~eV). A
time-resolved pump-probe ellipticity technique employing a
superconducting vector magnet allows us to measure all components of
the $g$-factor tensors, including their spread in the QD ensemble.
The $g$-factor anisotropy $|\delta g|$ reaches large values of about
1 for electrons and about 2.8 for holes. We show that the hole longitudinal $g$ factor
can be used to estimate key structural parameters of the QDs.

\section{Experimental details}

The samples under study were grown by molecular-beam epitaxy on a
(001)-oriented InP substrate. The QDs were formed by depositing 5.5
monolayers of InAs on a layer of In$_{0.53}$Al$_{0.24}$Ga$_{0.23}$As
also used for capping the dots \cite{Belykh2015}. The dot density is
about $10^{10}$~cm$^{-2}$. The diameter and height of the
optically active QDs are around 40-50 and 9-13~nm, respectively. Sample A
is nominally undoped, and sample B contains a Si $\delta$-doped
layer at a distance of 15~nm below the InAs QD layer. The samples
mostly differ in the QD emission energies. The QD
photoluminescence (PL) spectra of both samples consist of an
inhomogeneously broadened line of width $\sim 60$~meV, centered at
$\sim 0.81$~eV for sample A and at $\sim 0.78$~eV for sample B
\cite{Belykh2015}.

The samples are placed in a vector magnet system consisting of three
superconducting split-coils oriented orthogonally to each other. By
adjusting the currents in each coil the magnetic field magnitude (up
to 3~T) and direction can be varied. The samples are kept at
temperature $T=7-15$~K. A pump-probe technique with polarization
sensitivity is implemented to measure the carrier spin dynamics. As
laser source we use a NT\&C laser system consisting of an Optical
Parametric Amplifier (OPA) pumped by a mode-locked Yb:KGW laser
operating at 1040~nm \cite{Krauth2013}. The laser system generates a
periodic train (repetition rate $40$ MHz) of 300-fs-long pulses at
wavelengths tunable in the $1350-4500$~nm ($0.28-0.92$~eV) spectral
range. By a pulse shaper, the broad ($\sim 60$~nm) spectrum is
shaped down to a width of 20~nm (10~meV) centered at the desired
wavelength. Only $g$-factor energy dependence measurements were done for the laser spectrum shaped to a width of 10~nm (5~meV).
The average excitation power was about $10$~mW focused to a
spot of $50$~$\mu$m diameter, corresponding roughly to $\pi$-pulse
excitation as this leads to maximal carrier spin polarization. The
laser output is split into the pump and probe beams. The
circular-polarized pump pulses induce the carrier spin polarization,
whose temporal evolution is probed by measuring the ellipticity of
the probe beam, initially linearly polarized, after its transmission
through the sample. This method is analogous to measuring the
Faraday rotation of the probe beam and provides similar information
\cite{Glazov2010,Varwig2012}.

In addition, the population dynamics of the optically-injected
electron-hole pairs in the QDs is investigated by measuring the
differential transmission $\Delta \mathcal{T}/\mathcal{T}$ in a pump-probe experiment.
Linearly polarized pump pulses are used to generate the carrier
population that is monitored by the linearly polarized probe pulses
with variable delay relative to the pump pulses. Pump and probe
pulses have the same photon energy and orthogonal linear
polarizations to avoid  polarization interference.

\section{Results and discussion}

\begin{figure}
\begin{center}
\includegraphics[width=\columnwidth]{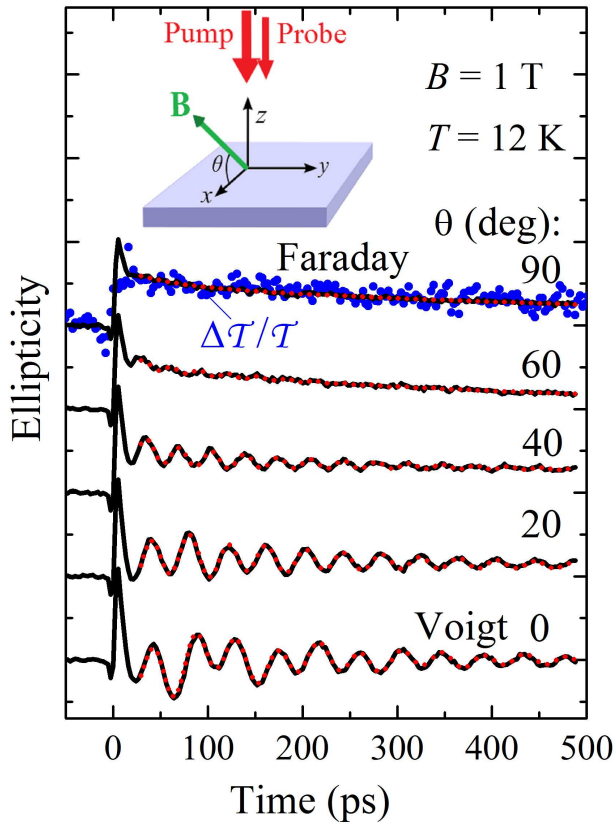}
\caption{Dynamics of the ellipticity signal at different angles
$\theta$ of the magnetic field $B=1$~T with respect to the sample
plane.  $\theta=0^\text{o}$ and $\theta=90^\text{o}$ correspond to
the Voigt and Faraday geometries, respectively. The curves are
shifted vertically for clarity. Red dotted lines show fits to the
experimental data. Blue solid circles show the dynamics of the
differential transmission at $B=0$~T. The data are shown for the
sample A at $T=12$~K. The central laser photon energy is set to 0.82~eV.
Inset shows the experimental geometry. } \label{FigKinXZ}
\end{center}
\end{figure}

The time dynamics of the spin polarization measured by the pump-probe
ellipticity for the magnetic field applied in Voigt geometry (along
the sample surface), shows the typical signatures of coherent spin
precession, namely, an oscillatory signal about zero level
\cite{Yakovlev2008} (see the lower curve in Fig.~\ref{FigKinXZ}). The
detailed analysis of this dynamics for sample B was presented in
Ref.~\cite{Belykh2015}, where it was shown that the oscillations
occur at two different frequencies, both of which show a linear
dependence on magnetic field. The high-frequency oscillations
correspond to the electron spin precession with $|g_{\text{e},x}|
\approx 1.6$ for sample A, while the low-frequency oscillations
correspond to the hole spin precession with $g$ factor
$|g_{\text{h},x}| \approx 0.6$.

We note, that in line with Ref.~\cite{Belykh2015}, the linear
dependencies of the oscillation frequencies on magnetic field show
no offset. According to Ref.~\cite{Yugova2007}, this indicates that an
exchange interaction of the electron and hole spins is small
compared to the Zeeman splitting, so that the oscillations occur on
the pure electron and hole spin precession frequencies, while
exciton effects can be neglected.

In the following, we describe the experiments in inclined magnetic
field using the coordinates $xyz$, where we chose the $z$ axis along
the sample growth direction (001), and the $x$ and $y$ axes in the
sample plane, so that the $x$ axis is either along the
crystallographic direction (110) or (1\={1}0). Note,
that in the experiment we detect the time evolution of the $z$
projection of the spin polarization, i.e., $S_z$.

\subsection{Out-of-plane anisotropy}

\begin{figure}
\begin{center}
\includegraphics[width=0.9\columnwidth]{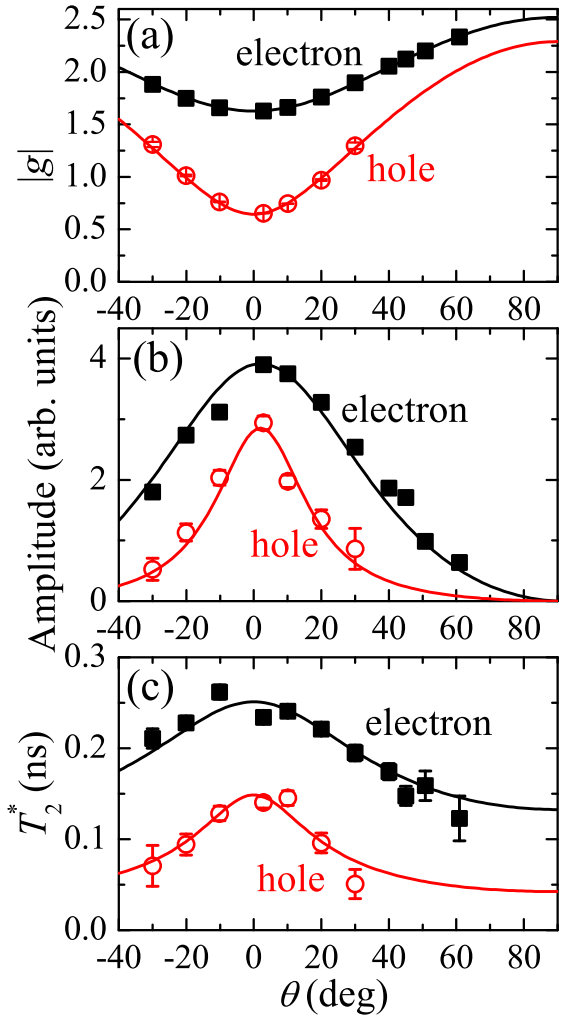}
\caption{Dependencies of the $g$-factor moduli (a), oscillation
amplitudes (b) and spin dephasing times (c) on the angle $\theta$
between the sample surface and the magnetic field for the electron
(solid squares) and hole (open circles) spin precession. Solid lines
in panels (a), (b), and (c) show fits to the experimental data with
Eqs.~\eqref{Eqgabs}, \eqref{EqSosc}, and \eqref{EqT2L}, respectively. The data are shown
for sample A at $B=1$~T and $T=12$~K. The laser photon
energy is set to 0.82~eV.} \label{FigThetaDep}
\end{center}
\end{figure}

For measurements of the out-of-plane anisotropy of carrier $g$
factors and spin dephasing times, the magnetic field direction was
first changed in the $xz$ plane (see inset of Fig.~\ref{FigKinXZ}).
The dynamics of the ellipticity signal at $B=1$~T for different
angles $\theta$ between $\mathbf{B}$ and the $x$ axis are shown in
Fig.~\ref{FigKinXZ}. When the angle $\theta$ is varied from
$0^\circ$ (Voigt geometry) to $90^\circ$ (Faraday geometry), the
precession frequencies increase, and the amplitude of the
oscillating signal becomes smaller up to its disappearance in
Faraday geometry. This is accompanied by the appearance of a
nonoscillating decaying component, which determines the signal in
the Faraday geometry. Also, the decay of the oscillating signal
becomes faster with increasing $\theta$. These variations are
summarized in quantitative form in Fig.~\ref{FigThetaDep} after
introduction of the fitting routine.

In the Faraday geometry, the spin dynamics shows only a
nonoscillating decay. In this case, the decay of the spin
polarization along the $z$ axis, and, correspondingly, along
$\mathbf{B}$ is determined by the spin lifetime $T_\text{S}$. This
time is contributed by the carrier recombination time,
$\tau_\text{rec}$, and the longitudinal spin relaxation time $T_1$:
$1/T_\text{S}=1/T_1+1/\tau_\text{rec}$. The $T_1$ time is usually
much longer than  the spin coherence time $T_2$ and the ensemble
spin dephasing time $T_2^*$ \cite{Kroutvar2004}. As one can see in
Fig.~\ref{FigKinXZ}, the dynamics of the ellipticity signal for
$\theta= 90^\text{o}$ coincides with the dynamics of the
photocarrier population with $\tau_\text{rec}\approx1$~ns measured
by the differential transmission experiment (the blue solid
circles). Therefore, we conclude that the ellipticity decay time
$T_\text{S}\approx 1$~ns is mainly determined by $\tau_\text{rec}$.
We note that in both samples we do not find indications
for a resident carrier population despite the Si-doping in sample
B.

\begin{figure}
\begin{center}
\includegraphics[width=0.6\columnwidth]{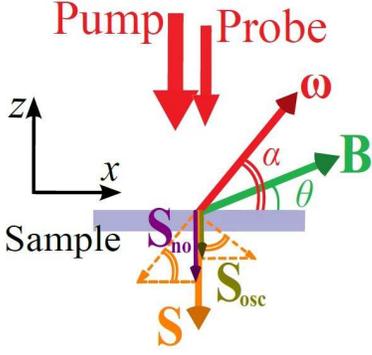}
\caption{Schematic representation of the vectors of magnetic field
$\mathbf{B}$, spin precession frequency $\boldsymbol{\omega}$ and
spin polarization $\mathbf{S}$. Note, that $\boldsymbol{\omega}$ is
not parallel to $\mathbf{B}$ due to the anisotropy of the $g$
factors. The spin polarization vector is decomposed into the
components rotating and nonrotating about $\boldsymbol{\omega}$. In
turn, the projections of these components on the probe beam
direction ($z$ axis) $\mathbf{S}_\text{osc}$ and
$\mathbf{S}_\text{no}$ determine the amplitudes of the oscillating
and nonoscillating components of the ellipticity signal.}
\label{FigBWS}
\end{center}
\end{figure}

The dynamics is fitted with the sum of two damped oscillating
functions of type $\cos(\omega t)\exp(-t/T_\text{2}^*)$, where $t$
is the time delay between the pump and probe pulses,
$\boldsymbol{\omega}=\hat{g}\mu_\text{B}\mathbf{B}/\hbar$ is the
Larmor precession frequency, and $\mu_\text{B}$ is the Bohr
magneton. Note, that the $T_\text{2}^*$ times measured in the
studied structures are typically shorter than 1~ns, i.e., the decay
of the coherent spin dynamics is mainly limited not by the carrier
recombination. To account for the nonoscillating component, a double
exponential decay was added to the fit. The dependencies of the
$g$-factor values, oscillation amplitudes and dephasing times on the
angle $\theta$ are shown in Fig.~\ref{FigThetaDep} for electron
(solid squares) and hole spins (open circles). The error bars
in the figure reflect the tolerated deviation of the parameters in
the fit of the dynamics. To describe the observed behavior of the
spin precession parameters we calculate the vector of the Larmor
precession frequency:
\begin{multline}
\left(\begin{array}{c}
    \omega_x \\
    \omega_y \\
    \omega_z
\end{array}\right) = \frac{\mu_\text{B}}{\hbar}
\left(\begin{array}{ccc}
    g_x & 0 & 0\\
    0 & g_y & 0 \\
    0 & 0 & g_z
\end{array}\right)
\left(\begin{array}{c}
    B\cos\theta\cos\varphi \\
    B\cos\theta\sin\varphi \\
    B\sin\theta
\end{array}\right)\\
=\frac{\mu_\text{B}B}{\hbar}\left(\begin{array}{c}
    g_x\cos\theta\cos\varphi \\
    g_y\cos\theta\sin\varphi \\
    g_z\sin\theta
\end{array}\right)
.
\label{EqWvec}
\end{multline}
Here $\varphi$ is the angle of the $\mathbf{B}$ projection on the
$xy$ plane with respect to the $x$ axis (see the inset in
Fig.~\ref{FigKinXY}(a)) and $g_x$, $g_y$ and $g_z$ are the diagonal
elements of the $g$-factor tensor in the selected basis of axes coinciding with the symmetry axes of the QDs.
Since the orientation of the magnetic field vector $\mathbf{B}$ is
varied in the $xz$ plane $\varphi=0$, and the $g$ factor modulus is
\begin{equation}
|g| = \frac{\hbar}{\mu_\text{B}B}\omega = \sqrt{(g_{x} \cos\theta)^2+(g_{z} \sin\theta)^2}.
\label{Eqgabs}
\end{equation}
The lines in Fig.~\ref{FigThetaDep}(a) show the fits to the
experimental data using Eq.~\eqref{Eqgabs} for electrons and holes.
From these fits we obtain the $x$ and $z$ components of the electron
$g$ factor, $|g_{\text{e},x}|=1.63$, and $|g_{\text{e},z}|=2.52$, as
well as the corresponding hole $g$ factors $|g_{\text{h},x}|=0.64$,
and $|g_{\text{h},z}|=2.29$.
%The entire $g$ factor anisotropy is shown in Fig.~\ref{FigPolar} in polar coordinates.

The dependence of the oscillation amplitude on $\theta$
[Fig.~\ref{FigThetaDep}(b)] can be modeled by calculating the
projection of the rotating spin component on the $z$ axis, as
illustrated in Fig.~\ref{FigBWS} \cite{Salis2001}. The optically
generated spin polarization is directed along the pump beam ($z$
axis). It rotates about $\boldsymbol{\omega}$, which is inclined
with respect to the sample plane $xy$ by the angle $\alpha =
\arctan(\omega_z/\omega_x)=\arctan(\tan\theta g_z/g_x)$ [see
Eq.~\eqref{EqWvec}]. So, the component of the spin polarization
$\mathbf{S}$ rotating about $\boldsymbol{\omega}$ is $S\cos\alpha$.
However, we detect only the projection of this rotating component on
the probe beam ($z$ axis) which is given by
\begin{equation}
S_\text{osc}=S\cos^2\alpha=\frac{S}{1+\tan^2\theta g_z^2/g_x^2}.
\label{EqSosc}
\end{equation}
The projection of the nonoscillating component on the probe
direction is
\begin{equation}
S_\text{no}=S\sin^2\alpha=\frac{S}{1+\cot^2\theta g_x^2/g_z^2}.
\label{EqScw}
\end{equation}
Note, that $S_\text{osc}+S_\text{no}=S$.

According to Eqs.~(\ref{EqSosc}) and (\ref{EqScw}) $S_\text{osc}$
decreases, while $S_\text{no}$ increases with $\theta$ in
agreement with the behavior shown in
Fig.~\ref{FigKinXZ}. Indeed, Eq.~(\ref{EqSosc}) gives a good fit to
the measured dependence of the oscillation amplitudes on $\theta$ as
shown in Fig.~\ref{FigThetaDep}(b). In the fits we use the $|g_x|$
and $|g_z|$ determined from the data in Fig.~\ref{FigThetaDep}(a).
It turns out that a better fit is obtained for $\theta$ shifted by
$\approx 2^\text{o}$, which is explained by the nonzero incidence
angle ($1^\text{o}-2^\text{o}$) of the probe beam with respect to
the growth axis, in order to suppress the influence of scattered
light from the degenerate pump.

To explain the dependence of the spin dephasing times $T_2^*$ on
$\theta$ [Fig.~\ref{FigThetaDep}(c)] we note that $T_2^*$ at
$B\gtrsim1$~T for these QDs is mostly determined by the
inhomogeneous spread, $\Delta g$, of the $g$ factor in the QD
ensemble \cite{Belykh2015}, especially since we use laser pulses
with 10-meV spectral width. The ellipticity signal is proportional
to $\int_{-\infty}^{\infty}\cos(g\mu_\text{B}Bt/\hbar)F(g-g_0)dg$,
where $F$ is the distribution function of the $g$ factor in the
optically excited ensemble. In the case of a Lorentzian
distribution, integration gives oscillations with an exponential
decay $\cos(g_0\mu_\text{B}Bt/\hbar)\exp(-t/T_2^*)$. In the case of
a Gaussian distribution, integration gives oscillations with a
Gaussian decay $\cos(g_0\mu_\text{B}Bt/\hbar)\exp(-t^2/2T_2^{*2})$.
Note, that for the studied samples both forms of the oscillating
function give similar parameters when fitted to the experimental
data \cite{Belykh2015}. The spin dephasing time is given by
\begin{equation}
\frac{1}{T_\text{2}^*}=\frac{\Delta g\mu_\text{B}B}{\hbar} ,
\label{EqT2}
\end{equation}
where the $g$-factor spread $\Delta g$ is equal to the half width at
half maximum (HWHM) of the Lorentzian distribution or the dispersion
of the Gaussian distribution. The $x$, $y$, and $z$ components of the
$g$-factor tensor have different spreads ($\Delta g_x$, $\Delta g_y$,
and $\Delta g_z$). The ellipticity signal for arbitrary magnetic
field orientation in the $xz$ plane is proportional to
$\int_{-\infty}^{\infty}\int_{-\infty}^{\infty}\cos\{[g_0+(g_x-g_{0x})\partial
g/\partial g_x+(g_z-g_{0z})\partial g/\partial
g_z]\mu_\text{B}Bt/\hbar\}F_x(g_x-g_{0x})F_z(g_z-g_{0z})dg_x dg_z$
leading to the decaying oscillatory forms described above, with
$T_\text{2}^*$ given by Eq.~\eqref{EqT2}, where the $g$-factor
spreads for the Lorentzian $g$-factor distributions $F_x$ and $F_z$
are:
\begin{equation}
\Delta g_\text{L}=\left|\frac{\partial g}{{\partial g_x}}
\right|\Delta g_x+\left|\frac{\partial g}{{\partial
g_z}}\right|\Delta g_z. \label{EqDgL}
\end{equation}
For the Gaussian distributions, these spreads are given by:
\begin{equation}
\Delta g_\text{G}=\sqrt{\left(\frac{\partial g}{{\partial
g_x}}\Delta g_x \right)^2+\left(\frac{\partial g}{{\partial
g_z}}\Delta g_z\right)^2}, \label{EqDgG}
\end{equation}
From Eqs.~\eqref{Eqgabs}, \eqref{EqT2} and
Eqs.~(\ref{EqDgL}-\ref{EqDgG}) one obtains the dephasing times for the Lorentzian and Gaussian distributions, respectively:
\begin{equation}
T_{2,\text{L}}^{*}=\frac{\hbar}{\mu_\text{B}B}\cdot\frac{\sqrt{g_{x}^2
\cos^2\theta+g_{z}^2 \sin^2\theta}}{|g_{x}|\Delta g_{x}
\cos^2\theta+|g_{z}| \Delta g_{z} \sin^2\theta}, \label{EqT2L}
\end{equation}
\begin{equation}
T_{2,\text{G}}^{*}=\frac{\hbar}{\mu_\text{B}B}\cdot\frac{\sqrt{g_{x}^2
\cos^2\theta+g_{z}^2 \sin^2\theta}}{\sqrt{(g_{x}\Delta g_{x}
\cos^2\theta)^2+(g_{z} \Delta g_{z} \sin^2\theta)^2}}. \label{EqT2G}
\end{equation}

The solid line in Fig.~\ref{FigThetaDep}(c) shows
fit to the experimental data with Eq.~\eqref{EqT2L}. In the fit we use the $|g_x|$
and $|g_z|$ determined from the data in Fig.~\ref{FigThetaDep}(a);
only $\Delta g_{x}$ and $\Delta g_{z}$ are taken as variables. We obtain the values $\Delta
g_{\text{e},x}=0.04$, $\Delta g_{\text{e},z}=0.09$ for the electron
and $\Delta g_{\text{h},x}=0.08$, $\Delta g_{\text{h},z}=0.27$ for
the hole spins. We have also determined the $T_2^{*}$ and fitted
$T_2^{*}(\theta)$ in the Gaussian approach (not
shown). The Gaussian approach gives slightly poorer agreement
(at least for electrons) compared to that in the Lorenzian approach, but
almost the same $\Delta g_{\text{e},x}$ and $\Delta g_{\text{h},x}$.
While the $\Delta g_{\text{e},z}$ and $\Delta g_{\text{h},z}$ are
about 1.2 times larger than the corresponding parameters for the
Lorentzian fit, corresponding to an even larger $\Delta g$
anisotropy.
%It is interesting to compare relative $g$ factor spreads which are $\Delta g_{\text{e}x}/g_{\text{e}x}=0.03$, $\Delta %g_{\text{e}z}/g_{\text{e}z}=0.09$, $\Delta g_{\text{h}x}/g_{\text{h}x}=0.04$, $\Delta g_{\text{h}z}/g_{\text{h}z}=0.18$.

\begin{figure}
\begin{center}
\includegraphics[width=0.9\columnwidth]{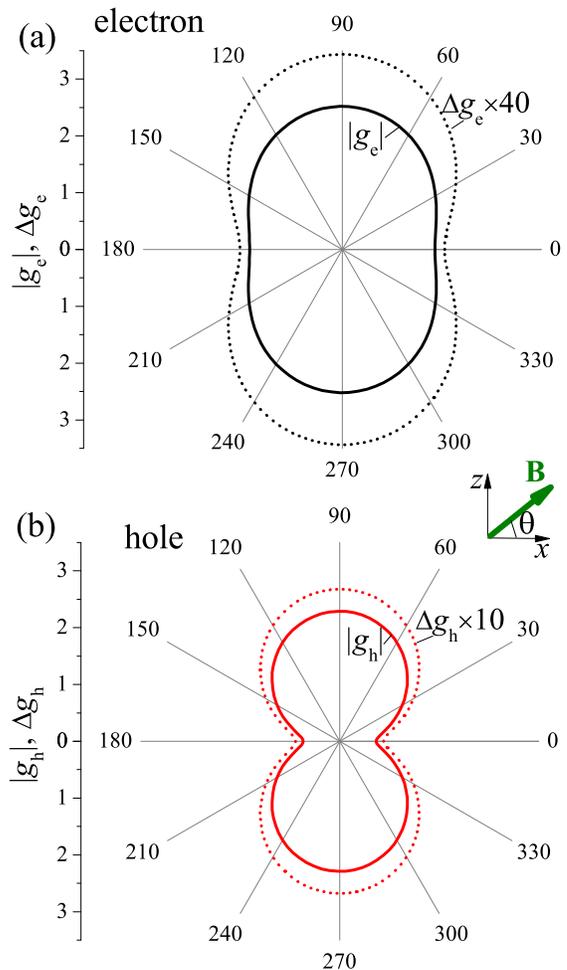}
\caption{Dependence of the $g$-factor moduli (solid lines) and their
spreads $\Delta g$ (dotted lines) for electrons (a) and holes (b) on
the angle $\theta$ between the sample surface and the magnetic
field. The dependencies are given by Eqs.~\eqref{Eqgabs} and
\eqref{EqDgL} parametrized with the $|g_x|$, $|g_z|$ and $\Delta
g_x$, $\Delta g_z$ that are determined from the experimental data
for sample A. The values of $\Delta g$ are multiplied by factors of
40 and 10 for electrons and holes, respectively, for better
visibility.} \label{Figgdg}
\end{center}
\end{figure}

Figure~\ref{Figgdg} compares the anisotropies of the electron and
hole $g$ factors given by Eq.~\eqref{Eqgabs} and their spread given
by Eq.~\eqref{EqDgL} in the Lorentzian approach for the parameters
determined from the fits in Fig.~\ref{FigThetaDep}. Interestingly,
the anisotropy of electron $\Delta g$ is somewhat larger than that of $g$.

\subsection{In-plane anisotropy}

\begin{figure}
\begin{center}
\includegraphics[width=0.9\columnwidth]{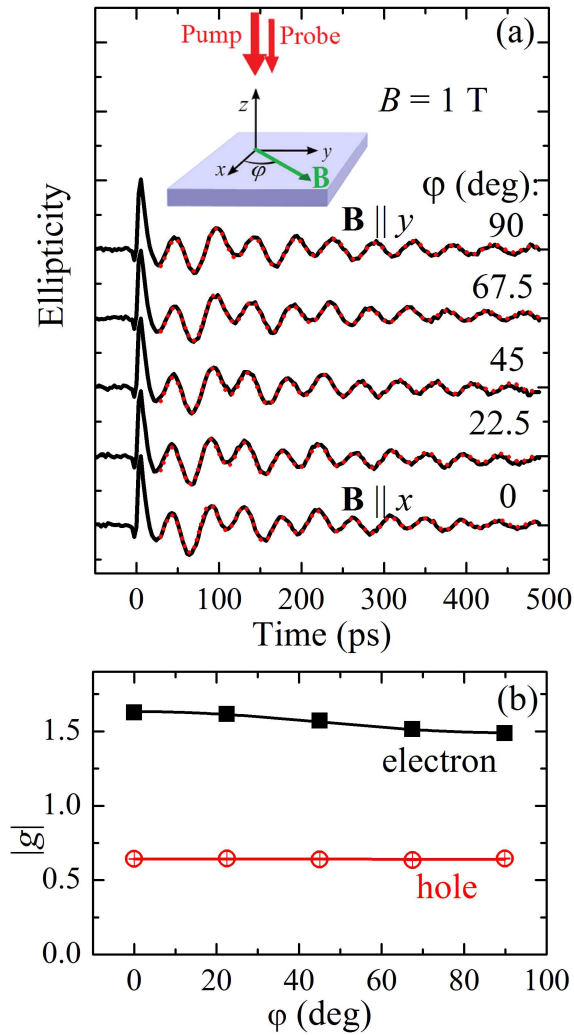}
\caption{(a) Dynamics of the ellipticity signal for the magnetic
field $\mathbf{B}$ orientation varied in the sample $xy$
plane, described by different angles $\varphi$. The curves are
shifted vertically for clarity. Red dotted lines show fits to the
experimental data. (b) Dependence of the $g$-factor moduli on the
angle $\varphi$ for the electrons (solid squares) and holes (open
circles). Solid lines show fits to the experimental data with
Eq.~\eqref{EqgabsXY}. The data are shown for the sample A at $B=1$~T
and $T=12$~K. The laser photon energy is set to 0.82~eV. Inset shows
the experimental geometry.} \label{FigKinXY}
\end{center}
\end{figure}

We also address the in-plane anisotropy of the spin precession by
changing the magnetic field vector $\mathbf{B}$ orientation in the
sample plane $xy$. Figure~\ref{FigKinXY}(a) shows the dynamics of
the ellipticity signal for different angles $\varphi$ of
$\mathbf{B}$ with respect to the $x$ axis. Varying $\varphi$ from
0$^\circ$ to 90$^\circ$ the amplitude and spin dephasing time remain
almost unchanged, while the precession frequency slightly decreases.
Figure~\ref{FigKinXY}(b) shows the $g$-factor dependencies on
$\varphi$ for electrons and holes. The solid lines show fits to the
experimental data with equation
\begin{equation}
|g| = \sqrt{(g_{x} \cos\varphi)^2+(g_{y} \sin\varphi)^2},
\label{EqgabsXY}
\end{equation}
obtained from Eq.~\eqref{EqWvec} for $\theta=0^\text{o}$. From the
fits we obtain $|g_{\text{e},x}|=1.63$, $|g_{\text{e},y}|=1.49$ for
the electrons and $|g_{\text{h},x}|=|g_{\text{h},y}|=0.64$ for the
holes. Surprisingly, for the holes with their complex band structure
the in-plane anisotropy vanishes within the experimental accuracy. The electrons on the other hand show a small in-plane anisotropy. From the spin
dephasing time at $\varphi = 90^\text{o}$ we evaluate the $g$-factor
spreads along the $y$ axis: $\Delta g_{\text{e},y}=0.04$ and $\Delta
g_{\text{h},y}=0.07$.

\subsection{Energy dependence}

\begin{figure}
\begin{center}
\includegraphics[width=0.9\columnwidth]{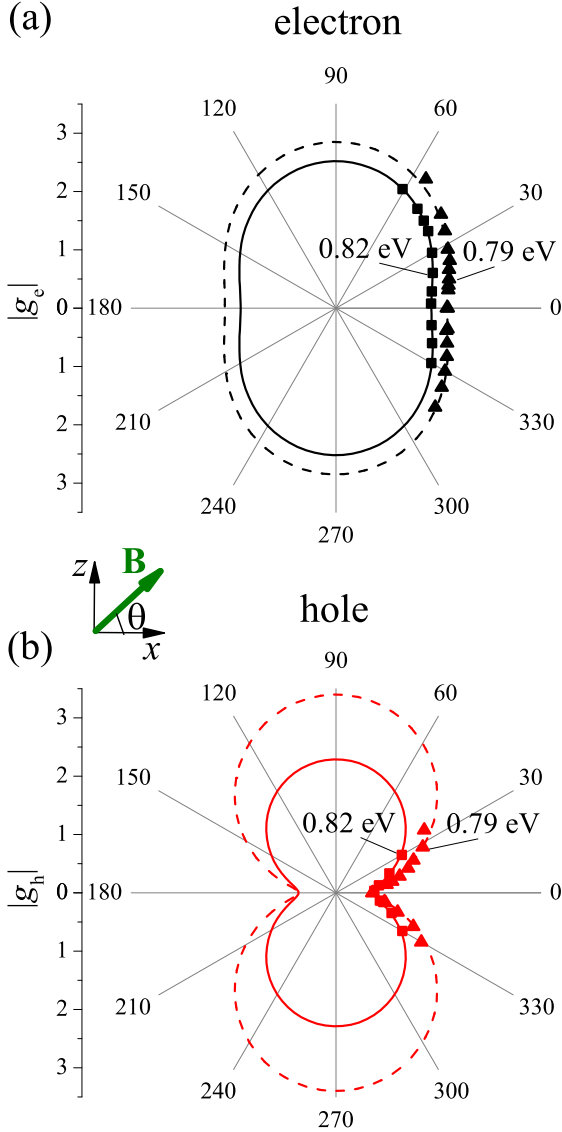}
\caption{Dependence of the $g$-factor moduli for electrons (a) and
holes (b) on the angle $\theta$ between the sample surface and the
magnetic field for the samples A (squares, 0.82~eV
transition energy) and B (triangles, 0.79~eV). Lines show fits to the experimental
data with Eq.~\eqref{Eqgabs}.} \label{FigPolar}
\end{center}
\end{figure}

We also studied the electron and hole spin precession for different
orientations of the magnetic field in the sample B with a laser
energy of 0.79~eV close to the mean energy of the ground-state transition. Figure~\ref{FigPolar} compares the out-of-plane
$g$-factor anisotropies for the two samples A and B. The
$g_\mathrm{e}$ anisotropy remains almost unchanged at the decreased
transition energy [Fig.~\ref{FigPolar}(a)]: The $g$ factor increases
by about $13\%$ for the in-plane direction and the direction normal
to it. On the other hand, the $g_\mathrm{h}$ anisotropy is strongly
enhanced with the energy decrease: $|g_{\text{h},x}|$ decreases
slightly, but $|g_{\text{h},z}|$ increases strongly from 2.29 up to
3.40. For the sample B, the in-plane $g$-factor anisotropy stays
small with $|g_{\text{e},x}|=1.90$, $|g_{\text{e},y}|=1.69$ and
$|g_{\text{h},x}|=0.63$, $|g_{\text{h},y}|=0.58$. The main
$g$-factor results for both samples are summarized in
Table~\ref{Tab1}.

\begin{table}
\begin{center}
\begin{tabular}{| c  |  c |  c |  c  |  c |}
    \hline
     & \multicolumn{2}{ |c| }{Electron} & \multicolumn{2}{ |c| }{Hole} \\
    %\hline
    \cline{2-5}
     &0.79 eV & 0.82 eV & 0.79 eV & 0.82 eV\\
    \hline
    $|g_x|$ & 1.90 & 1.63 & 0.63 & 0.64\\
    $|g_y|$ & 1.69 & 1.49 & 0.58 & 0.64\\
    $|g_z|$ & 2.85 & 2.52 & 3.40 & 2.29\\
    \hline
    $\Delta g_x$ & & 0.04 & & 0.08 \\
    $\Delta g_y$ & & 0.04 & & 0.07 \\
    $\Delta g_z$ & & 0.09 & & 0.27 \\
    \hline
    $T_2^*(B=1$~T$, \theta = 0^\circ)$ & &230~ps & &140~ps\\
    $T_2^*(B=1$~T$, \theta = 30^\circ)$ & &200~ps & &50~ps\\
    \hline
    \end{tabular}
\caption{Components of the electron and hole $g$-factor tensors and
their spreads along the $x$, $y$, and $z$ axes for the two
transition energies of 0.79 and 0.82~eV in the samples B and A, respectively. The spin dephasing times
at $B=1$~T for $\theta=0^\circ$ and $30^\circ$ are also given.}
\label{Tab1}
\end{center}
\end{table}

\begin{figure*}
\begin{center}
\includegraphics[width=1.8\columnwidth]{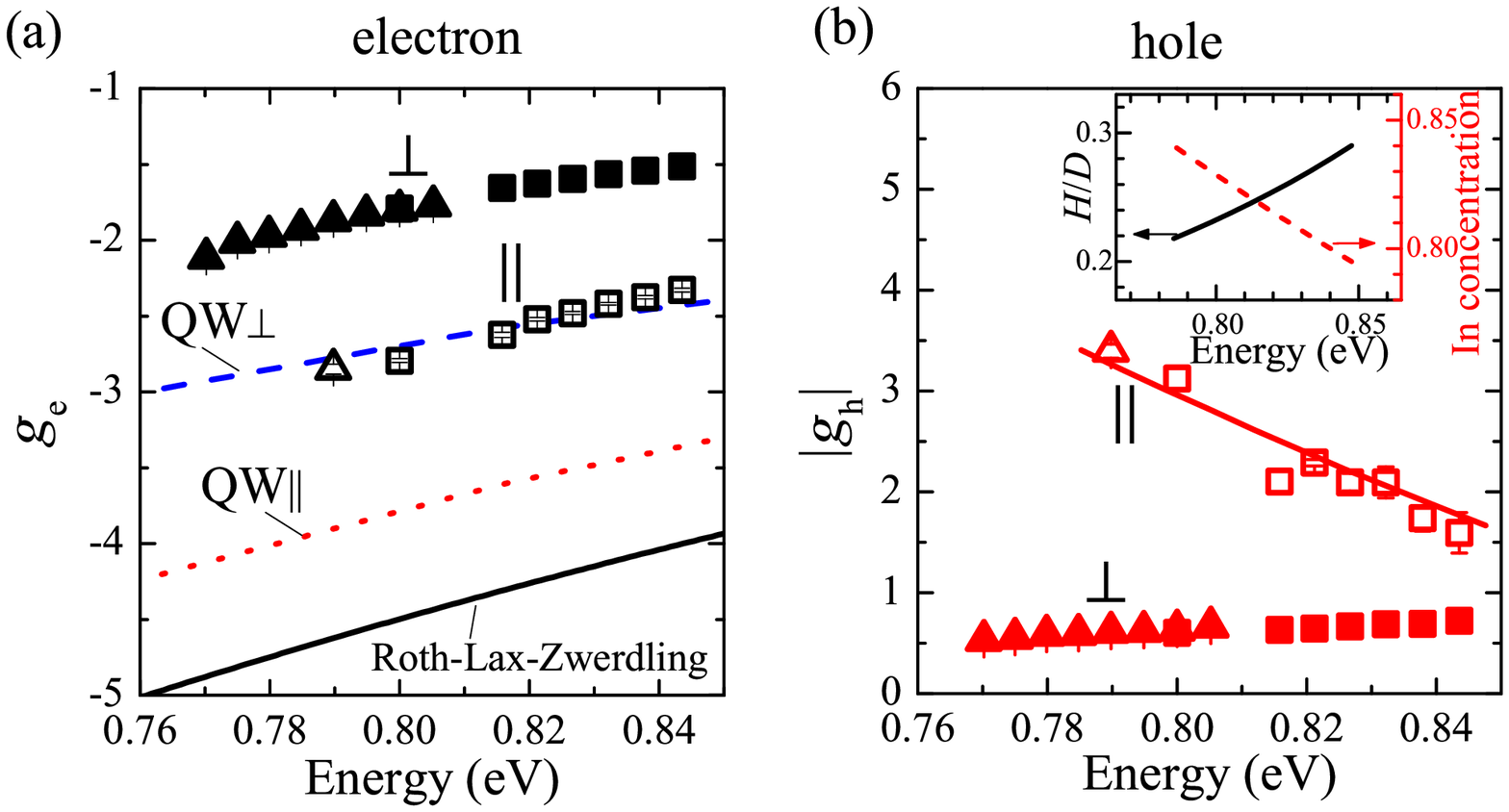}
\caption{Dependence of the electron (a) and hole (b) $g$ factors on
the central laser photon energy. The solid and open symbols show the
transverse and longitudinal $g$ factors, respectively. The data
corresponding to samples A and B are shown by squares and triangles,
respectively. The solid line in panel (a) shows the expected
dependence for bulk semiconductors calculated according to the
Roth-Lax-Zwerdling relation~\eqref{EqRLZ}. The dashed and
dotted lines in panel (a) show the transverse and
longitudinal $g$-factor dependencies, respectively,
calculated for QWs using the model of Ref.~\cite{Kiselev1998}.
In panel (b) the solid line shows the calculated longitudinal
hole $g$ factor using the model of Refs.~\cite{Semina2015,
Semina2016}. Inset shows the corresponding calculated dependencies
of the height-to-diameter ratio (solid line, left axis) and the
indium content (dashed line, right axis) on the QD emission
energy.} \label{FigEDep}
\end{center}
\end{figure*}

The energy dependencies of the electron and hole $g$ factors are
shown in Fig.~\ref{FigEDep}. The solid and open symbols show the transverse
($g_{\text{e},\perp} \equiv g_{\text{e},x}$) and longitudinal
($g_{\text{e},\parallel} \equiv g_{\text{e},z}$) $g$ factors,
respectively. The data shown by squares are taken from sample A, and
the data shown by triangles from sample B. We take the sign of the
electron $g$ factor to be negative, based on previous measurements
of the dynamic nuclear polarization in (In,Ga)As/GaAs QDs emitting
at larger energies \cite{Yugova2007}. The correctness of the
electron $g$-factor sign is further confirmed by the fact that both
transverse and longitudinal electron $g$ factors increase with
energy following the expected trend, see Fig.~\ref{FigEDep}(a):
Increase of the electron $g$ factor with energy was also reported
in a number of papers on QDs \cite{Greilich2006Sci, Schwan2011a}, in
agreement with the Roth-Lax-Zwerdling relation for bulk
semiconductors \cite{Roth1959}:
\begin{equation}
g_\text{e}(E_\text{g}) = g_\text{0} - \frac{2E_\text{p}
\Delta_\text{SO}}{3E_\text{g}(E_\text{g}+\Delta_\text{SO})}.
\label{EqRLZ}
\end{equation}
Corresponding calculations are shown in Fig.~\ref{FigEDep}(a) by the
solid line as function of the band gap energy $E_\text{g}$
\cite{Belykh2015}. As material parameters we use a spin-orbit
splitting $\Delta_\text{SO}=0.374$~eV of the valence band and a Kane energy
$E_\text{p}=24.0$~eV, obtained by linear interpolation between the
InAs and GaAs parameters using an average band gap energy $E_\text{g}=0.8$~eV. We
added $g_\text{remote}=-0.13$ to the calculated $g$ factor to
account for the contribution from the remote bands
\cite{Kiselev1998, Yugova2007Univ}.

\begin{table}
\begin{center}
\begin{tabular}{| c  || c | c || c | c |}
    \hline
        & $\Delta g_x$ & $\Delta g_{xE}$  & $\Delta g_z$ & $\Delta g_{zE}$ \\
    \hline
    Electron & 0.04 & 0.03 & 0.09 & 0.05 \\
    \hline
    Hole & 0.08 & 0.02 & 0.27 & 0.15\\
    \hline
    \end{tabular}
\caption{Comparison of the $g$-factor spreads $\Delta g$ along the $x$ and $z$ axes,
evaluated on one hand from the spin dephasing time and on the other
hand from the $g$-factor dispersion at 0.82~eV energy and laser spectral width $\Delta E=10$~meV for sample A.} \label{Tab2}
\end{center}
\end{table}

It is instructive to compare $\Delta g$ evaluated from the spin
dephasing time (Table~\ref{Tab1}) with $\Delta g_E=|\partial
g/\partial E| \Delta E /2$ from the $g$-factor dispersion $g(E)$ and
the spectral width of the laser $\Delta E=10$~meV used in the experiments where corresponding dephasing time was determined. The factor $1/2$
arises from the definition of $\Delta g$ as HWHM of the $g$-factor
distribution. This comparison is shown in Table~\ref{Tab2} for the
$x$ and $z$ directions. $\Delta g_E \gtrsim \Delta g/2$ except of
$\Delta g_{\text{h},x}$. Thus, the $g$-factor dispersion accounts
for a significant part of $\Delta g$. Indeed, measurements with a
smaller laser spectral width $\Delta E=5$~meV on sample A give
$\Delta g_{\text{e},x}\approx 0.03$ and $\Delta
g_{\text{h},x}\approx 0.07$, smaller than those measured for $\Delta
E=10$~meV.

Let us turn to the electron $g$-factor dispersions shown in
Fig.~\ref{FigEDep}(a). The large difference between the transverse
and longitudinal $g$ factors, $\delta g =
g_{\text{e}\perp}-g_{\text{e}\parallel} \sim 1$, indicates that at the
small band-gap energies $E_\text{g}$, for strong confinement, the electron $g$ factor is
strongly affected by the directional variations of the QDs, e.g., in strain and in size, as they have a diameter of $\sim
45$~nm and a height of $\sim 10$~nm. Indeed, electron $g$-factor anisotropy is dominated by the hole confinement, and, e.g., in QWs $\delta g$ is proportional to the difference of light- and heavy-hole energies \cite{Ivchenko2005,Sirenko1997}. Note, that in (In,Ga)As/GaAs QDs emitting around 1.4~eV and
having comparable height-to-diameter ratio, but much smaller confinement,
the  $g_\mathrm{e}$ anisotropy is much smaller ($\delta g_\mathrm{e}
\sim 0.1$) \cite{Yugova2007,Schwan2011}. Also,  small  is the
$g_\mathrm{e}$ deviation from the Roth-Lax-Zwerdling relation.

Our observation of a large electron $g$-factor anisotropy is
qualitatively confirmed by the model of Ref.~\cite{Kiselev1998}. The
dashed and dotted lines in Fig.~\ref{FigEDep}(a) show the
calculated energy dependencies of the transverse and
longitudinal electron $g$ factors, $g_{\text{e}\perp}$ and
$g_{\text{e}||}$, respectively, for a QW. The QW approximation of
the model \cite{Kiselev1998} is reasonable for the studied
dome-shaped QDs, since their diameter is much larger than their
height. The calculation details and used parameters can be found in
Refs.~\cite{Kiselev1998} and \cite{Belykh2015}, respectively. The calculations give approximately the same
difference $g_{\text{e}\perp}(E)-g_{\text{e}||}(E)$ as observed in
experiment. However, there is some deviation between experiment and
calculations in the absolute $g$-factor values. A more refined
theoretical description of electron $g$ factors taking into account
effects of strain, composition etc is still needed to reach
quantitative agreement with the experimental data.

We also calculate the longitudinal $g$ factor for holes,
$g_{\text{h}||}$, in the framework of the numerical method developed
in Refs.~\cite{Semina2015, Semina2016}. The quantum dots are modeled
as disks with Gaussian potential profiles for electrons and holes
$V_{e(h)}$:
\begin{equation}\label{anG}
V_{e(h)}(r,z)=V_{c(v)}\left(1-\exp\left[-\frac{4r^2}{D^2}-\frac{4z^2}{H^2}\right]\right).
\end{equation}
Here $V_{c(v)}$ is the conduction (valence) band discontinuity
between QD and barrier, $D$ and $H$ are the QD effective diameter
and height, respectively. The ratio  $V_{c}:V_{v}$ is taken as
$6:4$. The QD composition is assumed to be gradually varying between
the QD center, In$_p$Al$_q$Ga$_q$As, $p+2q=1$, and the barrier,
In$_{0.53}$Al$_{0.24}$Ga$_{0.23}$As. The dependence of the band gap
on the composition is taken from Ref.~\cite{Vurgaftman2001}. Other
band structure parameters are linearly interpolated between the
corresponding values of pure InAs, AlAs and GaAs also taken from
\cite{Vurgaftman2001}. Calculations show that the $g_{\text{h}||}$
is determined by the height-to-diameter ratio of the QDs, $H/D$, and
to less extent by the QD composition. On the other hand, the QD
emission energy is more strongly affected by the QD composition than
by the QD size. The solid line in Fig.~\ref{FigEDep}(b) shows the
calculated dependence for $H=11$~nm, $D$ varied from 50 to 38~nm,
and the Indium content in the QD center varied from 0.84 to 0.80.
The corresponding dependencies on the $H/D$ ratio (solid line, left
axis) and In content (dashed line, right axis) on the QD emission
energy are shown in the inset of Fig.~\ref{FigEDep}(b). Thus, with
increasing QD emission energy the $H/D$ increases and the In
concentration decreases.

\section{Conclusions}

By implementing a vector magnet into a setup for time-resolved
pump-probe ellipticity studies, we performed a detailed study of the
$g$-factor anisotropy of electrons and holes in
InAs/In$_{0.53}$Al$_{0.24}$Ga$_{0.23}$As self-assembled quantum
dots, emitting in the telecom spectral range around 0.8~eV. All
components of the $g$-factor tensors were measured, as well as their
spreads in the QD ensemble. The electron $g$ factor shows a
comparatively huge out-of-plane anisotropy changing from $g_{\mathrm{e},x}=
-1.63$ to $g_{\mathrm{e},z}= -2.52$ at a transition energy of
0.82~eV. The hole $g$-factor anisotropy at this energy is even
stronger: $|g_{\text{h},x}|= 0.64$ and $|g_{\text{h},z}|= 2.29$. It
increases even further at a smaller transition energy of 0.79~eV:
$|g_{\text{h},x}|= 0.63$ and $|g_{\text{h},z}|= 3.40$. The spread of
the $g$ factors determined from the spin dephasing time shows a
pronounced out-of-plane anisotropy, both for electrons and holes. On
the other hand, the in-plane anisotropy of the electron and hole $g$
factors is small. The
energy dependence of the longitudinal hole $g$ factors has been
described applying the theoretical model of Refs.~\cite{Semina2015,
Semina2016}, which has allowed us to estimate the QD parameters:
size and composition. In particular, it is shown that with
increasing emission energy, the height-to-diameter ratio of the QDs
increases while the indium content in QD decreases.

\begin{acknowledgements}
We are grateful to A.~Greilich, E.~L.~Ivchenko, A.~A.~Kiselev and
I.~A.~Yugova for valuable advices and useful discussions. We
acknowledge the financial support by the Russian Science Foundation (Grant No.
14-42-00015), the Deutsche Forschungsgemeinschaft in the frame of the ICRC TRR 160, as well as by BMBF in the
frame of the project Q.com-H (Contracts No. 16KIS0112 and No. 16KIS0104K).
\end{acknowledgements}
\end{document}